\documentstyle[preprint,epsfig,aps,floats]{revtex}
\begin{document}
\draft
\preprint{\vbox{
\hbox{IFT-P.024/99}
\hbox{March 1999}
}}
\title{ L-R asymmetries and signals for new bosons~\thanks{Presented 
by M. C. Rodriguez at VIII Mexican School of Particles and Fields, Oaxaca de 
 Ju\'arez, Oax, M\'exico, November 20--29, 1998. 
}}
\author{
J. C. Montero~\thanks{e-mail: montero@ift.unesp.br}, V. Pleitez~\thanks{
e-mail: vicente@ift.unesp.br} and M. C. Rodriguez~\thanks{e-mails: 
mcr@axp.ift.unesp.br}
}
\address{
Instituto de F\'\i sica  Te\'orica\\ 
Universidade  Estadual Paulista\\
Rua Pamplona, 145\\ 
01405-900-- S\~ao Paulo, SP\\ 
Brazil } 
\maketitle
\begin{abstract}
Several left-right parity violating asymmetries in lepton-lepton scattering 
in fixed target and collider experiments are considered as signals for
doubly charged vector bosons (bileptons).  
\end{abstract}
\newpage
The left-right asymmetry when only one of the lepton is polarized is 
defined as follows~\cite{dm}
\begin{equation}
A_{RL}(ll\to ll)=\frac{d\sigma_R-d\sigma_L}{d\sigma_R+d\sigma_L},
\label{asy1}
\end{equation}
where $d\sigma_{R(L)}$ is the differential cross section for one right
(left)-handed lepton $l$ scattering on an unpolarized lepton $l$.
That is
\begin{equation}
A_{RL}(ll\to ll)=\frac{(d\sigma_{RR}+d\sigma_{RL})-(d\sigma_{LL}
+d\sigma_{LR})}{(d\sigma_{RR}+d\sigma_{RL})+(d\sigma_{LL}
+d\sigma_{LR})},
\label{asy2}
\end{equation}
where $d\sigma_{ij}$ denotes the cross section for incoming leptons with
helicity $i$ and $j$, respectively, and they are given by 
\begin{equation}
d\sigma_{ij}
\propto\sum_{kl}\vert M_{ij;kl}\vert^2,\quad i,j;k,l=L,R.
\label{dsigma}
\end{equation}
Another interesting possibility is the case when both leptons are 
polarized. We can define an asymmetry $A_{R;RL}$ in which
one beam is always in the same polarization state, say right-handed, and 
the other  is either
right- or left-handed polarized (similarly we can define  $A_{L;LR}$):
 \begin{equation}
A_{R;RL}=\frac{d\sigma_{RR}-d\sigma_{RL}}{d\sigma_{RR}+d\sigma_{RL}},\qquad
A_{L;RL}=\frac{d\sigma_{LR}-d\sigma_{LL}}{d\sigma_{LL}+d\sigma_{LR}}.\qquad
\label{ar}
\end{equation}

We can define also an asymmetry when one incident particle is right-
handed and the other is left-handed and the final states are right- and left or
left- and right-handed:
\begin{equation}
A_{RL;RL,LR}=\frac{d\sigma_{RL;RL}-d\sigma_{RL;LR}}{d\sigma_{RL;RL} 
+ d\sigma_{RL;LR}}
\label{ar3}
\end{equation}
or similarly, $A_{LR;RL,LR}$. These asymmetries, in Eqs.(\ref{ar}) and 
(\ref{ar3}), are dominated by QED 
contributions. However, this will not be the case if a bilepton resonance
does exist at typical energies of the NLC. To show this fact is the goal of 
this paper.
These asymmetries can be calculated for both fixed target (FT) and 
collider (CO) experiments.

We can integrate in the scattering angle and define the asymmetry 
$\overline{A}_{RL}$ as
\begin{equation}
\overline{A}_{RL}(ll\to ll)=\frac{(\int d\sigma_{RR}+\int d\sigma_{RL})-(\int 
d\sigma_{LL}
+\int d\sigma_{LR})}{(\int d\sigma_{RR}+\int d\sigma_{RL})+(\int d\sigma_{LL}
+\int d\sigma_{LR})},
\label{aint}
\end{equation}
where $\int d\sigma_{ij}\equiv \int^{175^o}_{5^o}d\sigma_{ij}$.
All these asymmetries can be studied in future accelerators~\cite{nlc,gunion}.


The importance of these sort of parity breaking asymmetries in fixed target 
experiments in lepton-lepton scattering was first pointed out in 
Ref.~\cite{dm}. For the case of electron-electron scattering the mass of the 
electrons can be neglected. For an energy
of $E=50$ GeV and for a scattering angle (in the center of mass frame) of 
$\theta=90^o$ the left-right asymmetry, defined in 
Eq.~(\ref{asy1}) has a value $\approx-3\times10^{-7}$ in the standard model.
Radiative corrections reduce this value about $40\pm3\,\%$~\cite{cm}. 
It is expected that fixed target experiments like those at SLAC~\cite{slac}
can measure this asymmetry~\cite{cm}.
For the muon-muon elastic scattering this asymmetry is $\approx5.4
\times10^{-5}$~\cite{mpr1}. We have studied also the non-diagonal elastic
scattering $e\mu\to e\mu$. In the last case we obtain a value of 
$-5.9\times10^{-8}$ for a muon energy of 50 GeV and $-2.9\times10^{-7}$ for 
muon energy of 190 GeV. At these energies the muon mass cannot be 
neglected~\cite{mpr2}.
This type of asymmetry can be measured using the high-energy muon beam M2 of 
the CERN SPS as in the NA47 experiment~\cite{smc}.  


The relevance of these asymmetries in collider experiments was first pointed 
out in Refs.~\cite{mpr1,cm2}. In fixed target experiments the cross sections 
are large ($\sim \mbox{mb}$) and the asymmetries small ($\sim10^{-7}$). On the
other hand, in collider experiments the cross sections are small 
($\sim10^{-3}\mbox{nb}$) but the asymmetries large ($\sim0.1$ for the 
muon-muon case). 
Explicitly we have that at energies $\sqrt{s}=300$ GeV and $\theta\approx90^o$
the asymmetry is 
\begin{equation}
A^{\rm CO, ESM}_{RL}(ee\to ee)\approx -0.05,
\label{asec}
\end{equation}
for the electron-electron case and
\begin{equation}
A^{\rm CO,ESM}_{LR}(\mu\mu\to\mu\mu)\approx-0.1436,
\label{asmc}
\end{equation}
for the muon-muon case.
Future colliders with polarized lepton-lepton scattering can have the 
appropriate luminosity to measure these parameters.    

If a muon-electron collider is constructed in the future, it would be possible
to measure the  $A_{RL}^{CO;ESM}(\mu e)=-0.024$ for $E_\mu=190$ GeV 
($\sqrt{s}\sim 380$ GeV) and $\theta=90^o$. At high energies the mass effects 
are not important.


So far all the results were obtained in the 
standard model. In certain kind of models there are doubly charged scalars
($H^{--}$) or/and vector ($U^{--}$) bileptons~\cite{331}.   
As expect the asymmetry is larger in the $U$-pole. For instance,
\begin{equation}
A^{\rm CO,ESM+U}_{RL}(ee\to ee)= -0.099,
\label{asec331}
\end{equation}
and
\begin{equation}
A^{\rm CO,ESM+U}_{RL}(\mu\mu\to\mu\mu)= -0.1801,
\label{asmcu}
\end{equation}
when we add to the standard model asymmetry the 
contributions due to the the bilepton $U$ for $M_U=300$ GeV and $\Gamma_U=36$
MeV. In Fig.~1 we show the behaviour of
the asymmetry $A^{\rm CO,ESM+U}_{RL}$ as a function of the mass of the
boson $U^{--}$.

For the electron-electron case we can define the quantity
\begin{equation}
 \delta\, \overline{A}_{RL}(ee\to ee)\equiv 
(\overline{A}\,^{\rm CO,ESM+U}_{RL}-\overline{A}\,^{\rm CO,ESM}_{RL})/
\overline{A}\,^{\rm CO,ESM}_{RL},
\label{deltadef}
\end{equation} 
where $\bar{A}$'s are define in Eq.~(\ref{aint}). Although 
$\delta\, \overline{A}_{RL}$ is large (near 50 for $\sqrt{s}=300$) at the 
$U$-resonance we would like to 
stress that it remains appreciably large even far from the $U$-peak. That
particular behavior suggests that this quantity could be the one to be 
considered in the search for new physics, like the bilepton $U^{--}$, in 
future colliders. 
On the other hand, the asymmetry is insensitive to the 
contributions of the doubly charged scalars. 

We have used also the asymmetries defined in Eq.~(\ref{ar}). In this case
it is interesting to note that
\begin{equation}
A^{\rm CO,ESM+U}_{R;RL}(ee\to ee)\approx -A^{\rm CO,ESM}_{R;RL}(ee\to ee),
\label{f1}
\end{equation}
and we see that such a difference on sign is a good signature for the
discovery of the vector bilepton.

The contributions of an extra neutral vector boson $Z'$ has also been 
considered for the case $A_{RL}(\mu e)$. 
In this case the asymmetry is considerably enhanced and it will
be appropriate in searching for extra neutral vector bosons with mass up to
1 TeV. Since in the 331 model the $Z'$ couplings with the leptons are
flavor conserving we do not have additional suppression factors coming from 
mixing~\cite{331}. Hence, the $\mu e$ elastic scattering can be very helpful, 
even with the present experimental capabilities, for looking for non-standard 
physics effects.

\acknowledgments 
This work was supported by Funda\c{c}\~ao de Amparo \`a Pesquisa
do Estado de S\~ao Paulo (FAPESP), Conselho Nacional de 
Ci\^encia e Tecnologia (CNPq) and by Programa de Apoio a
N\'ucleos de Excel\^encia (PRONEX).

\vglue 0.01cm
\begin{figure}[ht]
\begin{center}
\vglue -0.01cm
\mbox{\epsfig{file=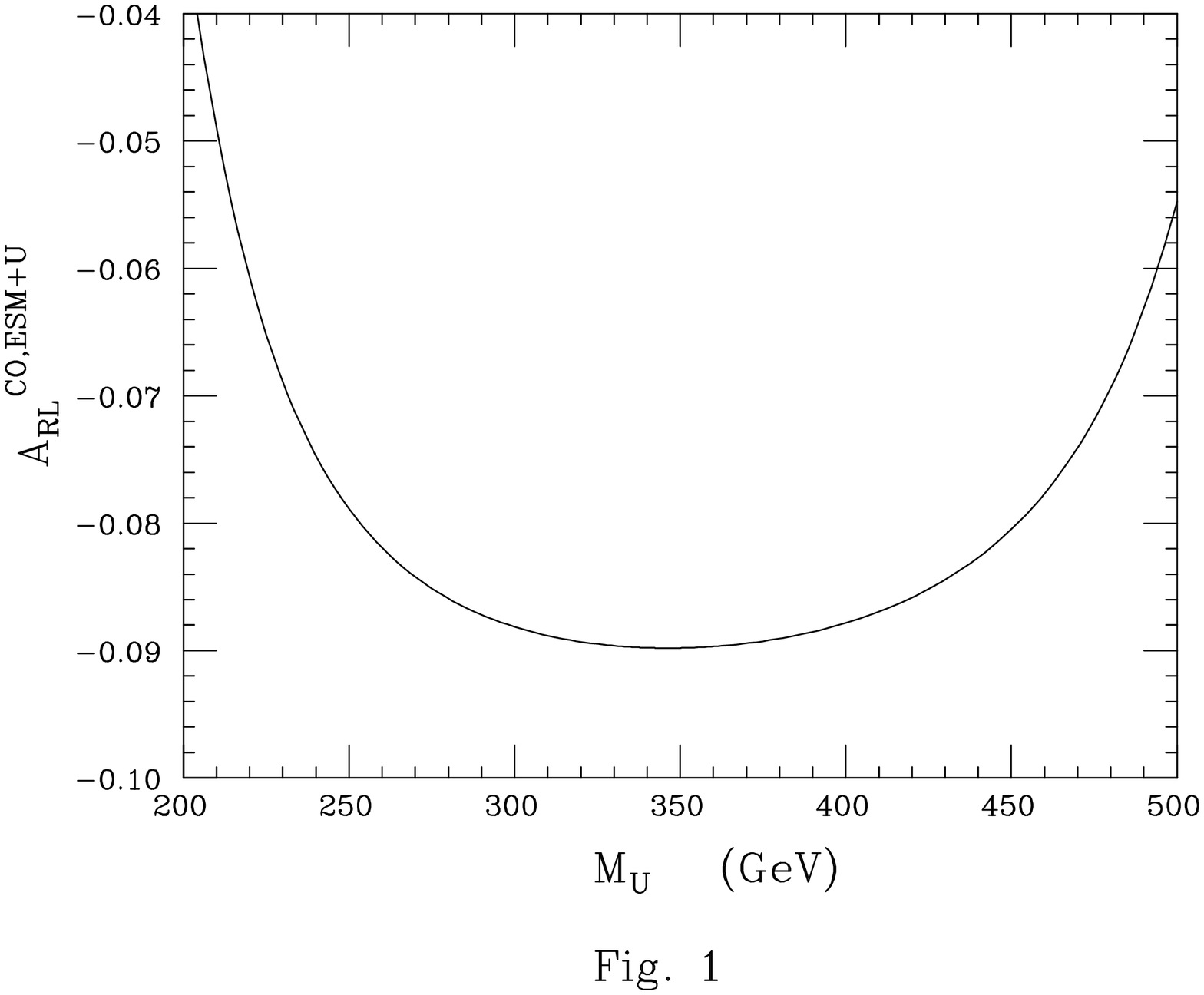,width=0.7\textwidth,angle=90}}       
\end{center}
\caption{ $L-R$ asymmetry as a function of the $M_U$ mass for 
$\theta=\pi/2$, $\sqrt{s}=300$ GeV and $\Gamma_U=36$ MeV.}
\label{fig1}
\end{figure}

\end{document}